\documentclass[12pt]{article}

\usepackage{amsfonts,amssymb,amsthm,amsmath}
\usepackage[dvips]{graphics}

\advance\textheight\topskip
\newcommand{\mysection}[1]{\section{#1}\setcounter{equation}{0}}

\newcommand{\vdd}[2]{{\frac{\delta #1}{\delta #2}}}
\def\bea{\begin{eqnarray}}
\def\eea{\end{eqnarray}}

\newcommand{\bref}[1]{\textbf{\ref{#1}}}

\renewcommand{\d}{\partial}

\renewcommand{\geq}{\,{\geqslant}\,}
\renewcommand{\leq}{\,{\leqslant}\,}

\newcommand{\binner}[2]{%
  {\langle}\kern-4.15pt{\langle}#1{,}\,#2{\rangle}\kern-4.15pt{\rangle}}

\newcommand{\half}{\mathchoice{%
    \ffrac{1}{2}}{\frac{1}{2}}{\frac{1}{2}}{\frac{1}{2}}}

\newcommand{\ffrac}[2]{\raisebox{.5pt}%
  {\footnotesize$\displaystyle\frac{#1}{#2}$}\kern1pt}

\def\cA{\mathcal{A}}

\def\cC{\mathcal{C}}

\def\cE{\mathcal{E}}

\def\cI{\mathcal{I}}
\def\cJ{\mathcal{J}}

\def\cL{\mathcal{L}}

\def\cS{\mathcal{S}}
\def\cT{\mathcal{T}}

\def\5{\bar }
\def\6{\partial }
\def\7{\hat }
\def\4{\tilde }

\def\Q#1#2{\frac{\partial #1}{\partial #2}}

\def\eps{\epsilon}

\def\dx{\text{d}\hspace{-0.06em}x}

\begin{document}
\pagestyle{myheadings}
\markboth{\textsc{\small Barnich, Comp\`ere}}{%
  \textsc{\small Geometric derivation of Smarr relation for Kerr AdS
    BH}} \addtolength{\headsep}{4pt}

%% \begin{titlepage}

\begin{flushright}
  ULB-TH/04-31\\
  \texttt{gr-qc/0412029}
\end{flushright}

\begin{centering}

  \vspace{0.5cm}

  \textbf{\Large{Generalized Smarr relation
    for Kerr AdS black \\ \vspace{10pt}
     holes from improved surface integrals}}

  \vspace{1.5cm}

  {\large G.~Barnich$^{*}$ and G.~Comp\`ere$^\dag$}

  \vspace{1.5cm}

\begin{minipage}{.9\textwidth}\small \it \begin{center}
    Physique Th\'eorique et Math\'ematique \\ Universit\'e Libre de
    Bruxelles\\and\\ International Solvay Institutes\\ Campus
    Plaine C.P. 231, B-1050 Bruxelles\\ Belgium \end{center}
\end{minipage}

\end{centering}

\vspace{1cm}

\begin{center}
  \begin{minipage}{.9\textwidth}
    \begin{center} \textbf{Abstract}\\\vspace{10pt}\end{center}
    By using suitably improved surface integrals, we give a unified
    geometric derivation of the generalized Smarr relation for higher
    dimensional Kerr black holes which is valid both in flat and in
    anti-de Sitter backgrounds. The improvement of the surface
    integrals, which allows one to use them simultaneously at infinity
    and on the horizon, consists in integrating them along a path in
    solution space. Path independence of the improved charges is 
    discussed and explicitly proved for the higher dimensional Kerr
    AdS black holes. It is also shown that the charges for these black
    holes can be correctly computed from the standard Hamiltonian or
    Lagrangian surface integrals. 
  \end{minipage}
\end{center}

\vfill

\noindent
\mbox{}
\raisebox{-3\baselineskip}{%
  \parbox{\textwidth}{ \mbox{}\hrulefill\\[-4pt]}} {\scriptsize$^*$
  Senior Research Associate of the National
  Fund for Scientific Research (Belgium). E-mail: gbarnich@ulb.ac.be\\[-2pt]
  $^\dag$ Research Fellow of the National Fund for Scientific
  Research (Belgium). E-mail: gcompere@ulb.ac.be}

\thispagestyle{empty} \newpage

\tableofcontents

\mysection{Introduction}

The geometric derivation of the Smarr relation and of the first law
of thermodynamics for four dimensional asymptotically flat black holes
is usually based on Komar integrals \cite{Bardeen:1973gs,Carter:1972}.
Even though they do not provide a complete and systematic approach to
conserved quantities\footnote{In order to give the correct definitions
  of energy and angular momentum, the coefficients of the Komar
  integrals must be fixed by comparison with the ADM expressions
  \cite{Arnowitt:1962aa,ADM} (see e.g.\cite{Townsend:1997ku}).}, Komar
integrals are extremely useful since they allow one to easily express the
conserved quantities defined at infinity to properties associated with the
horizon of the black hole. This approach can be extended to higher
dimensional asymptotically flat black holes \cite{Myers:1986un}, but
generally fails or becomes rather cumbersome for rotating asymptotically
Anti-De Sitter black holes.

As has been emphasized recently \cite{Gibbons:2004ai}, not even in
four dimensions do all authors obtain the same expression for the
energy of Kerr AdS black holes and some of these expressions are in
disagreement with the first law. Gibbons et al.~compute the energy of
such black holes indirectly by integrating the first law. In
\cite{Deruelle:2004mv}, the mass and energy have been computed
directly by using the BKL superpotentials \cite{Katz:1996nr}. In a
completed version of their paper, Gibbons et al.~then have also
computed the energy directly by using the Ashtekar-Magnon-Das
definition \cite{Ashtekar:1984,Ashtekar:1999jx}.

In this article, we first briefly discuss the standard surface
integrals at infinity that are used to define the conserved charges
associated to Killing vectors. Several approaches to getting these
surface integrals exist: a direct approach by Abbott and Deser based
on manipulating the linearized Einstein equations
\cite{Abbott:1982ff}, the Hamiltonian approach
\cite{Regge:1974zd,Henneaux:1985tv,Henneaux:1985ey}, covariant phase
space methods \cite{Iyer:1994ys,Wald:1999wa}, covariant Noether
methods \cite{Julia:1998ys,Silva:1998ii,Julia:2000er} and
cohomological techniques \cite{Anderson:1996sc,Barnich:2001jy}. We
will recall various expressions that one obtains and their
relations\footnote{We will not discuss the quasi-local approach
  \cite{Brown:1993br}, which has been used in the present context in 4
  dimensions in \cite{Caldarelli:1999xj}.}.

We then recall that the surface integrals can be improved by
integrating them along a path in solution space \cite{Wald:1999wa}.
Like the Komar integrands, the improved integrands are closed
\cite{Barnich:2003xg} wherever the matter-free Einstein equations
hold, so that Stokes' theorem can be used and the conserved quantities
do not depend on the surfaces used for their evaluation. In
particular, the conserved quantities computed over the $n-2$-sphere at
infinity can be expressed as integrals over any other surface which,
together with the $n-2$ sphere at infinity, bound an $n-1$ dimensional
hypersurface.

The original part of the paper starts with a detailed discussion of
the integrability conditions that guarantee that the charges computed
with the improved surface integrals do not depend on the path used for
the improvement, but only on the background solution and the end point
solution.

We then compute the conserved charges, mass and angular momenta, for
the Kerr AdS black holes by using the improved surface integrals and
find agreement with the the results of
\cite{Gibbons:2004ai,Deruelle:2004mv}.  We also show explicitly that,
in this case, the improved surface integrals reduce to the standard
Lagrangian or Hamiltonian surface integrals at infinity, which thus
also allow one to correctly compute the charges and, at the same
time, proves the path independence of the improved charges. 

Finally, we give a detailed and geometric derivation of the
generalized Smarr relation for the higher dimensional Kerr AdS black
holes, as outlined in \cite{Barnich:2003xg}. The derivation can also
be applied to asymptotically flat black holes. We do not need to do
this explicitly as the corresponding results are recovered
straightforwardly in the limit of vanishing cosmological constant.

\mysection{Conserved quantities at infinity}

\subsection{Covariant expressions} \label{covexp}

A systematic approach to surface integrals in general relativity
consists in classifying all conserved $n-2$ forms, i.e., all $n-2$
forms built out of the metric and a finite number of their derivatives
such that the exterior derivative vanishes for all solutions of
Einstein's equations. One finds \cite{Barnich:1995db,Barnich:2000zw}
that all such $n-2$ forms are given by forms which vanish on all
solutions up to exterior derivatives of $n-3$ forms. As a consequence,
the associated charges obtained by integrating these forms for a given
solution over closed $n-2$ dimensional surfaces all vanish.

If one considers the same problem for linearized general relativity
around a fixed background solution, one finds that the non trivial
conserved $n-2$ forms are in one-to-one correspondence with the
Killing vectors $\bar \xi$ of the background $\bar g$
\cite{Barnich:1995db,Anderson:1996sc,Barnich:2004ts}. The
corresponding surface integrals should then be used only at a
boundary, where the deviations $h_{\mu\nu}$ from the background are
small and the linearized approximation is justified (see
\cite{Barnich:2001jy} for more details). In the following, we will
have in mind the case where this boundary is $S^\infty$, the $n-2$
sphere at infinity. The charges are then given by
\begin{eqnarray}
  \label{eq:19}
  Q^\infty_{\bar\xi}=\oint_{S^\infty} k_{\bar\xi}[h;\bar g].
\end{eqnarray}

Explicitly, the $n-2$ forms $k_\xi[h,\bar g]$ can be obtained from the
Killing vectors $\bar \xi^\mu$ through so-called descent equations.
One finds\footnote{For convenience, the conserved $n-2$ forms have
  been defined with an overall minus sign as compared to the
  definition used in \cite{Barnich:2001jy}, and the Killing vectors of
  the background metric are denoted by $\bar \xi^\mu$ instead of
  $\tilde\xi^\mu$. Finally, as compared to the definition used in
  \cite{Barnich:2003xg}, we will include an overall factor of
  $\frac{1}{16\pi}$ in the definition of the Komar integrand
  $K^K_{\bar \xi}[g]$ below.}
\begin{eqnarray}
 k_{\bar
\xi}[h;\bar g]=\frac{1}{16\pi}(d^{n-2}x)_{\mu\nu} \sqrt{- \bar g}\Big(
\bar\xi^\nu \bar D^\mu h+\bar \xi^\mu \bar D_\sigma h^{\sigma\nu}+\bar
\xi_\sigma \bar D^\nu h^{\sigma\mu}\nonumber\\+\half h\bar
D^\nu\bar\xi^\mu+\half h^{\mu\sigma}\bar D_\sigma \bar\xi^\nu+\half
h^{\nu\sigma} \bar D^\mu \bar\xi_\sigma -(\mu\longleftrightarrow \nu)
\Big), \label{hom}
\end{eqnarray}
where indices are lowered and raised with the background metric $\bar
g_{\mu\nu}$ and its inverse, $\bar D_\mu$ and $h_{\mu\nu}$ denote,
respectively, the covariant derivative and the deviation with respect to
this background metric. We use here and in the following
\begin{eqnarray}
  \label{eq:1}
  (d^{n-p}x)_{\mu_1\dots\mu_p}&\equiv&
\frac 1{p!(n-p)!}\, \epsilon_{\mu_1\dots\mu_n}
dx^{\mu_{p+1}}\dots dx^{\mu_n},\quad
\epsilon_{0\dots (n-1)}=1,\\
d\sigma_i&\equiv& 2(d^{n-2}x)_{0i},\quad h=g^{\mu\nu}h_{\mu\nu}.
\end{eqnarray}
This expression can be shown to coincide with the one derived by
Abbott and Deser \cite{Abbott:1982ff}:
\begin{eqnarray}
k_{\bar \xi}[h;\bar g]=-\frac{1}{16\pi}(d^{n-2}x)_{\mu\nu} \sqrt{- \bar
g}\Big(\bar \xi_\rho \bar D_\sigma H^{\rho\sigma\mu\nu} +\frac 12
H^{\rho\sigma\mu\nu} \bar D_\rho\bar \xi_\sigma \Big), \label{gsuperpot3}
\end{eqnarray}
where $H^{\rho\sigma\mu\nu}[h;g]$ is defined by
\begin{eqnarray}
H^{\mu\alpha\nu\beta}[h;g]&=& -\7h^{\alpha\beta}\bar g^{\mu\nu}
-\7h^{\mu\nu}\bar g^{\alpha\beta} +\7h^{\alpha\nu}\bar g^{\mu\beta}
+\7h^{\mu\beta}\bar g^{\alpha\nu},
\label{Hdef}\\
\7h_{\mu\nu}&=& h_{\mu\nu}-\frac{1}{2} \bar g_{\mu\nu} h. \label{hath}
\end{eqnarray}

Using the exact Killing equation $\bar D_\mu\bar\xi_\nu+\bar
D_\nu\bar\xi_\mu=0$, one can simplify\footnote{We consider here and in
  the following only exact Killing vectors and not asymptotic ones,
  for which such simplifications require more care (cf. classical
  central extensions \cite{Brown:1986nw}).} (\ref{hom}) to the
expression derived in \cite{Anderson:1996sc}:
\begin{eqnarray}
  \label{eq:and}
  k_{\bar \xi}[h;\bar g]=
\frac{1}{16\pi}(d^{n-2}x)_{\mu\nu} \sqrt{-\bar g}\Big( h^{\mu\sigma}\bar
D_{\sigma}\bar\xi^\nu-\bar \xi_\sigma \bar D^\mu h^{\nu\sigma}- \half h
\bar D^\mu\bar \xi^\nu \nonumber\\+\bar \xi^\mu (\bar D_\sigma
h^{\nu\sigma}-\bar D^\nu h)-(\mu\longleftrightarrow
\nu)\Big).\label{anderson}
\end{eqnarray}
If $h_{\mu\nu}=\delta g_{\mu\nu}$, this last expression can be written
as
\begin{eqnarray}
  \label{eq:wald}
  k_{\bar \xi}[h;\bar g]=
-\delta (K^K_{\bar \xi}[g]) + K^K_{\delta \bar \xi
}[g]-\bar\xi\cdot\Theta[h;g]
\end{eqnarray}
where
\begin{eqnarray}
K^K_{\bar\xi}[g]=(d^{n-2}x)_{\mu\nu}
\frac{\sqrt{-g}}{16\pi}\Big(D^\mu\bar\xi^\nu-(\mu\leftrightarrow \nu)\Big)
\end{eqnarray}
is the Komar integrand,
\begin{eqnarray}
\Theta[h;\bar g]=(d^{n-1}x)_\mu\frac{\sqrt{-\bar
  g}}{16\pi}\Big(\bar D_\sigma
h^{\mu\sigma}-\bar D^\mu h\Big),
\end{eqnarray}
$\bar\xi\cdot=\bar\xi^\nu \frac{\partial}{\partial(dx^\nu)}$ is the
inner product and $\delta \bar \xi$ is defined such that $\cL_{\delta
  \bar \xi} \bar g_{\mu\nu} +\cL_{\bar \xi} \delta g_{\mu\nu}= 0$.  In
the case where the Killing vectors of the background and of the
perturbed solution $g_{\mu\nu}=\bar g_{\mu\nu}+\delta g_{\mu\nu}$ are
the same, $\delta\bar\xi^\mu=0$, $\cL_{\bar \xi} \delta g_{\mu\nu}=0$,
expression \eqref{eq:wald} coincides with the expression derived in
\cite{Iyer:1994ys}\footnote{A geometric derivation of the first law,
  based on \eqref{eq:wald} and valid without additional assumptions
  on the nature of the variation, will be presented elsewhere.}.

Finally, if $\delta\bar\xi=0$, the expression derived in
\cite{Katz:1996nr}
\begin{eqnarray}
  \label{eq:BKL}
  k^{BKL}_{\bar \xi}[g;\bar
  g]&=&-K^K_{\bar\xi}[g]+K^K_{\bar\xi}[\bar g]
%\nonumber\\&&
-(d^{n-2}x)_{\mu\nu}\frac{\sqrt{-g}}{16\pi}\Big(\bar\xi^\mu
k^\nu-(\mu\leftrightarrow \nu)\Big),
\end{eqnarray}
where
\begin{eqnarray}
  \label{eq:BKLbis}
k^\nu=g^{\nu\rho}(\Gamma^\sigma_{\rho\sigma}-\bar
\Gamma^\sigma_{\rho\sigma})-g^{\rho\sigma}(\Gamma^\nu_{\rho\sigma}-
\bar\Gamma^\nu_{\rho\sigma}),
\end{eqnarray}
coincides to first order in $h_{\mu\nu}$ with \eqref{eq:wald} as can
easily be seen by using $\delta \Gamma^\nu_{\rho\sigma}=\frac 12 (\bar
D_\sigma h^\nu_\rho+\bar D_\rho h^\nu_\sigma-\bar D^\nu
h_{\rho\sigma})$. Hence, this expression will give the same results if
evaluated at infinity and if the boundary conditions are such that the
terms quadratic and higher in $h_{\mu\nu}$ vanish asymptotically.

\subsection{Hamiltonian expression}
Starting from the action of general relativity in first order
Hamiltonian form and applying the general construction of conserved
$n-2$ forms of the linearized theory along the lines of
\cite{Barnich:2001jy}, one can write the $n-2$ form related to the Killing
vector $\bar \xi^\mu$, $\mu = 0,i$ of the background as
\begin{eqnarray}
k_{\bar\xi}[\delta\gamma,\delta \pi;\bar\gamma,\bar\pi]|_{x^0=cte} =
\frac{1}{16\pi} (d\sigma)_{a} \Big( \bar G^{abcd}\big[\bar\xi^\perp\bar \nabla_b
\delta\gamma_{cd} 
-\bar \nabla_b\bar\xi^\perp \delta\gamma_{cd}\big]\nonumber\\
 + 2\bar\xi^b \delta
\pi_b^{\,\,\,a} - \bar\xi^a \delta\gamma_{cd} \bar \pi^{cd} \Big),
\label{eq:k0iHamilt}\\
\bar G^{abcd} =\half \sqrt{\bar \gamma}(\bar \gamma^{ac}\bar \gamma^{bd} +
\bar \gamma^{ad}\bar \gamma^{bc} - 2 \bar \gamma^{ab}\bar \gamma^{cd}),
\end{eqnarray}
where $x^0$ has been assumed to be constant and $\delta \pi_b^{\,\,\,
  a} = \delta (\gamma_{ac}\pi^{cb})$ is understood.  In this
expression, $a=1,\dots ,n-1$, $\bar \gamma_{ab}$ denotes the spatial
background three metric, which is used, together with its inverse
$\bar \gamma^{bc}$ to lower and raise indices, $\bar \nabla_a$ is the
associated covariant derivative, $\bar \pi^{ab}$ are the conjugate
momenta, $\bar \xi_a=\delta_a^i\bar \xi_i$, with $i=1,\dots n-1$ and
$\bar \xi^\perp=N\bar \xi^0$, with $N$ the lapse function. This
expression coincides in the case of asymptotically anti-de Sitter
spacetimes with the expression derived in
\cite{Henneaux:1985tv,Henneaux:1985ey}.  General results on the
relation between the Hamiltonian and the Lagrangian formalism (see
e.g.  \cite{Henneaux:1992ig}) then imply that the charges computed in
the two approaches coincide.

\mysection{Improved surface integrals}

\subsection{Integrating along a path}

So far, in order to compute the charges the idea was to ``go to
infinity and stay there'' \cite{Julia:2000sy}. What allows one to ``go
into the bulk'' is the following modification of the $n-2$ forms
$k_{\bar\xi}[h,\bar g]$: consider a path $\gamma$ in the space of
solutions $g_{\mu\nu}$ to Einstein's equations that interpolates
between the background $\bar g_{\mu\nu}$ and a given solution
$g_{\mu\nu}$ and let $\xi^\mu_\gamma$ be a Killing vector field for
all the metrics along this path. Let $d_V g_{\mu\nu}$ denote a 1-form
on the space of metrics (see
e.g.~\cite{Andersonbook,Anderson1991,Olver:1993} for more details).
The $n-2$ form in coordinate space
\begin{eqnarray}
  \label{eq:4}
  K_{\xi;\gamma}=\int_\gamma k_{\xi_\gamma}[d_Vg;g]
\end{eqnarray}
obtained by integrating $k_{\xi_\gamma}[d_Vg;g]$, which is a 1-form in
field space, along the path $\gamma$ can be shown\footnote{In
  \cite{Barnich:2003xg}, only the case where the Killing vector $\bar
  \xi$ is the same along the whole path was explicitly considered. The
  extension of the proof to the case of a path dependent Killing
  vector fields that we will need for some of the computations below
  is straightforward.} to be closed wherever the interpolation is
meaningful,
\begin{eqnarray}
  \label{eq:8}
  dK_{\xi;\gamma}=0,
\end{eqnarray}
unlike the $n-2$ forms $k_{\bar\xi}[h,\bar g]$, which are merely closed
``at infinity''.

Explicitly, let $g^{(s)}_{\mu\nu}$ with $s\in [0,1]$ denote a one
parameter family of solutions to Einstein's equations interpolating
between the background $\bar g_{\mu\nu}=g^{(0)}_{\mu\nu}$ and the given
solution $g_{\mu\nu}=g^{(1)}_{\mu\nu}$, let $\xi^s$ be a Killing vector
field for this family, $\cL_{\xi^{(s)}}g^{(s)}_{\mu\nu}=0$,
$\xi^{(0)}=\bar\xi$, $\xi^{(1)}=\xi$, and
$h^{(s)}_{\mu\nu}=\frac{d}{ds}g^{(s)}_{\mu\nu}$ be the tangent vector
to $g^{(s)}_{\mu\nu}$ in solution space. We have
\begin{eqnarray}
  \label{eq:7}
  K_{\xi;\gamma}=\int_0^1 ds\, k_{\xi^{(s)}}[h^{(s)};g^{(s)}].
\end{eqnarray}
It follows from Stokes theorem that the charges
\begin{eqnarray}
  \label{eq:10}
  Q_{\xi;\gamma}=\oint_S K_{\xi;\gamma}
\end{eqnarray}
do not depend on the closed $n-2$ dimensional hypersurface $S$ used
for their evaluation\footnote{This is the case as long as two such
  surfaces $S$ and $S^\prime$ are the boundary of an $n-1$ dimensional
  hypersurface $\Sigma$ where \eqref{eq:8} holds and where there are
  no singularities. We always assume this in the following.}
\begin{eqnarray}
  \label{eq:20}
  \oint_S K_{\xi;\gamma}=\oint_{S^\prime} K_{\xi;\gamma}.
\end{eqnarray}

\begin{figure}[htbp]
\begin{center}
\resizebox{0.35\textwidth}{!}{\mbox{\includegraphics{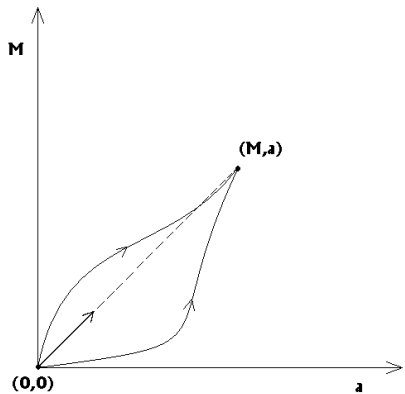}}}
\caption{\small{In the example of the four-dimensional Kerr black
    holes, the solution space is parameterized by the mass $M$ and
    rotation parameter $a$. One can for instance use the diagonal path
  $sM,sa$, $s\in[0,1]$ for the evaluation of the charges $Q_\xi$.}}
\label{integrability}
\end{center}
\end{figure}

\subsection{Path independence}\label{pathind}

The natural questions to ask for the charges $Q_{\xi;\gamma}$ are
whether they depend on the path $\gamma$ used in their definition and
what their relation with the charges $Q^\infty_\xi$ defined at
infinity is.

Suppose that the dependence on $s$ of $g^s_{\mu\nu}(x)$ and $\xi^s$ is
analytical and that in an expansion according to $s$ all terms which
are of order $s$ or higher vanish when one approaches the boundary at
infinity, 
\begin{eqnarray}
K_{\xi;\gamma}\longrightarrow k_{\tilde\xi}[h,\bar g],\label{Kk}
\end{eqnarray}
with $h=h^{(0)}$. Because the charges $Q_{\xi;\gamma}$ can be
evaluated on the surface $S^\infty$ at infinity, one finds, under this
assumption, that they agree with the charges defined at infinity,
$Q_{\xi;\gamma}=\oint_{S^\infty} k_{\tilde\xi}[h,\bar
g]=Q^\infty_\xi$. Furthermore, the charges $Q_{\xi;\gamma}$ then do
not depend on the path, but only on the initial solution $\bar
g_{\mu\nu}(x)$ and the final solution $g_{\mu\nu}(x)$.
This follows because, when evaluated at $S^\infty$, the charges are
manifestly path independent since only the initial solution and the
tangent vector pointing towards the final solution is involved. Since
furthermore, the charges do not depend on the surface used for their
evaluation, this remains true when they are evaluated at other
surfaces $S$ in the bulk.

For the Kerr AdS black holes considered below, the angular momenta will
be manifestly path independent, while we will show that the mass is
integrable because \eqref{Kk} holds.

Alternatively, in order to investigate path independence of the charges
$Q_{\xi;\gamma}$, one can study the integrability conditions
\cite{Wald:1999wa,Julia:2002df}
\begin{eqnarray}
  \label{eq:18}
  \oint_S d_V k_\xi[d_Vg;g]=0.
\end{eqnarray}
In the appendix, we will show that it follows directly from the
construction of the integrands $k_\xi[d_Vg;g]$ that the weak 
integrability conditions
\begin{eqnarray}
  \label{eq:17}
  d (d_Vk_\xi[d_Vg;g])|_{\delta g,\delta \xi;g,\xi}=0
\end{eqnarray}
hold when $d_Vk_\xi[d_Vg;g]$ is evaluated
for any $g_{\mu\nu},\delta_1 g_{\mu\nu},\delta_2
g_{\mu\nu}$, $\xi^\mu$, $\delta_1\xi^\mu,\delta_2\xi^\mu$ such that
\begin{enumerate}
\item $g_{\mu\nu}(x)$ is a solution to Einstein's equations,
\item $\delta g_{\mu\nu}(x)$ is a solution to the
linearized Einstein's equations,
\item $\xi^\mu(x)$ is a Killing vector for $g_{\mu\nu}(x)$,
\item $\delta\xi^\mu(x)$ satisfies the linearized Killing equation
$\cL_\xi \delta g_{\mu\nu}+\cL_{\delta\xi}
  g_{\mu\nu}=0$.
\end{enumerate}
Furthermore, we will also show that if there is no De Rham cohomology
in degree $2$ in solution space and no De Rham cohomology in degree
$n-2$ in $x^\mu$, the integrability conditions \eqref{eq:17} do indeed
guarantee integrability, i.e., path independence of the charges
$Q_{\xi;\gamma}$. 

It turns out however that, since the charges are
usually integrated over closed $n-2$ dimensional surfaces, there is
precisely one non vanishing De Rham cohomology class $c^{n-2}(x,dx)$
in form degree $n-2$. In the absence of De Rham cohomology in degree
$2$ in solution space, this class represents the only obstruction for
the weak integrability conditions \eqref{eq:17} to guarantee
conditions \eqref{eq:18} and thus path independence of the charges: 
$(d_Vk_\xi[d_Vg;g])|_{\delta g,\delta \xi;g,\xi}=k c^{n-2}+d(\cdot)$,
with $k$ a two form in solution space. 

In the following, we will assume that there is no such obstruction
($k=0$) and denote the path independent charges simply by $Q_\xi$. It
should be kept in mind however, that the charge related to the
solution $g_{\mu\nu}$ with Killing vector $\xi$ is measured with
respect to the background solution $\bar g_{\mu\nu}$ with Killing
vector $\bar \xi$.

Note that any of the equivalent expressions \eqref{hom},
\eqref{gsuperpot3}, \eqref{anderson}, \eqref{eq:wald} (or
\eqref{eq:k0iHamilt}, if $x^0$ is constant) can be used to define the
charges at infinity and also the improved charges defined in
\eqref{eq:7}. For convenience, we shall use expression
\eqref{eq:wald}, because the interpretation of the various terms will
be particularly simple. Indeed, in this case, one gets
\begin{eqnarray}
  \label{eq:3}
  Q_\xi= -\oint_S K^K_\xi[g]+\oint_S
  K^K_{\bar \xi}[\bar g]+\oint_S\cC_{\xi;\gamma},\\
\cC_{\xi;\gamma} = \int_0^1 ds
 \left( K^K_{\frac{d}{ds}\xi^s}[g^{(s)}]- \xi^{(s)} \cdot
\Theta[h^{(s)};g^{(s)}] \right), \label{intcC}
\end{eqnarray}
and the subtracted Komar integral appears explicitly. Note also that since
both $Q_\xi$ and $-\oint_S K^K_\xi[g]+\oint_S
  K^K_{\bar \xi}[\bar g]$ are path independent, so is
  $\oint_S\cC_{\xi;\gamma}$.

\mysection{Geometric derivation of generalized Smarr relation}

\subsection{Generalized Smarr relation}

We shall now show that the generalized Smarr relation follows directly
by evaluating the identity \eqref{eq:20}, where $S=S^\infty$, the
$n-2$ sphere at infinity and $S^\prime$ is chosen as an $n-2$
dimensional surface $H$ on the horizon.

More precisely, consider a black hole $g_{\mu\nu}$ with Killing horizon
determined by $\xi=k+\Omega_a m^a$, where $k$ denotes the time-like
Killing vector, $\Omega_a$ the angular velocities and $m^a$ the axial
Killing vectors.  The total energy of spacetime is defined to be $\cE
\equiv Q_k$, while the total angular momenta are $\cJ^a \equiv-Q_{m^a}$.
In the particular case where $S^\infty$ is chosen tangent to $m^a$ and if
the Killing vectors $m^a$ do not vary along the path $\gamma$, equation
(\ref{eq:3}) reduces to the standard expression for the angular momenta in
terms of Komar integrals:
\begin{eqnarray}
  \label{eq:6}
  \cJ^a=\oint_{S^\infty}
  K^K_{m^a}[g]-\oint_{S^\infty}
  K^K_{m^a}[\bar g].
\end{eqnarray}
For other surfaces or if the Killing vectors $m^a$ vary, the angular momenta
have to be computed with the more general expression (\ref{eq:3}), or,
if appropriate boundary conditions are fulfilled, with the expressions
\eqref{hom}, \eqref{gsuperpot3}, \eqref{anderson}, \eqref{eq:wald} (or
\eqref{eq:k0iHamilt} if $x^0$ is constant on $S^\infty$).

Suppose that there exists a path $g^{(s)}_{\mu\nu}$ in solution space
interpolating between the background $g^{(0)}_{\mu\nu}=\bar
g_{\mu\nu}$ and the black hole $g^{(1)}_{\mu\nu}=g_{\mu\nu}$ with
$k_{(s)}$ and $m^a_{(s)}$ Killing vectors along the whole path
reducing to $k,m^a$ at the end point $g_{\mu\nu}$ of the path. It is
not assumed that there is a horizon defined for all the metrics along
the path. The Killing vector $\xi^{(s)}$ along the path of metrics is
chosen as $\xi^{(s)}=k^{(s)}+\Omega^{(1)}_am^a_{(s)}$, where
$\Omega^{(1)}_a\equiv\Omega_a$ are the angular velocities of the
horizon of the final black hole described by $g_{\mu\nu}$. Note that
the Killing vector $\xi^{(s)}$ is then not the generator of the
Killing horizon for the metrics $g_{\mu\nu}^{(s)}$ even when those
metrics do describe black holes.

We have
\begin{eqnarray}
  \label{eq:5}
  \cE=\oint_{S^\infty}K_{k;\gamma}
  &=&\oint_{S^\infty} K_{\xi-\Omega_a
    m^a;\gamma}=\oint_HK_{\xi;\gamma}+\Omega_a\cJ^a.
\end{eqnarray}
The Komar integral $\oint_H K^K_\xi[g]$ evaluated on the horizon
is well known \cite{Bardeen:1973gs} to give $\frac{\kappa}{8\pi}\cA$ (see
also e.g.~\cite{Townsend:1997ku}), where $\cA$ is the area of the horizon.
We thus get
\begin{eqnarray}
  \label{eq:11}
  \cE-\Omega_a\cJ^a=\frac{\kappa}{8\pi}\cA+ \oint_H
    K^K_\xi[\bar g] + \oint_H \cC_{\xi;\gamma}.
\label{PreSmarr}
\end{eqnarray}
The claim is that this relation gives the generalized Smarr formula,
which becomes the thermodynamical Euler relation, with the standard
identifications of temperature as $\cT=\frac{\kappa}{2\pi}$ and
entropy as $\cS=\frac{1}{4}\cA$.

\subsection{The four-dimensional Schwarzschild black hole}

In order to gain some experience, let us quickly discuss the
four-dimensional Schwarz\-schild black hole, where $\Omega_a=0$ and
$\xi = k$. The only path interpolating between the flat background and
a given solution can be parameterized by replacing $M$ by $sM$,
$s\in[0,1]$, in the Schwarzschild solution in standard spherical
coordinates. It is then straightforward to verify that \eqref{Kk}
holds and $\cE=M$.  Furthermore, the second term on the right hand
side does not contribute for the flat background $\bar
g_{\mu\nu}=\eta_{\mu\nu}$. We then can compute $\oint_H\cC_\xi$ in two
ways:
\begin{enumerate}
\item Since the Komar integral is conserved by itself, $dK^K_\xi=0$,
  we also have $d\cC_\xi=0$. We thus can move back out to infinity:
  $\oint_H \cC_\xi=\oint_{S^\infty} \cC_\xi$. There, we can either
  compute $\cC_\xi$ directly or use the following arguments: (i) in
  standard spherical coordinates $\xi=\frac{\partial}{\partial t}$ is
  constant, so that $\cC_{\frac{\partial}{\partial t}}$ reduces to the
  $\Theta$ term, (ii) because of the fall-off conditions, higher order
  terms in $s$ do not contribute at infinity so that
  $\oint_{S^\infty}\cC_{\frac{\partial}{\partial t}}=
  -\oint_{S^\infty} \frac{\partial}{\partial t} \cdot \Theta[h;\bar g]$,
  (iii) the total mass at infinity is given by
  $-\oint_{S^\infty}(K^K_{\frac{\partial}{\partial t}}+
  \frac{\partial}{\partial t} \cdot \Theta[h;\bar g])=M$, (see
  e.g.~\cite{Iyer:1994ys}), (iv) since
  $-\oint_{S^\infty}K^K_{\frac{\partial}{\partial t}}=\half M$ (see
  e.g.~\cite{Bardeen:1973gs}) it follows that
  $\oint_{S^\infty}\cC_{\frac{\partial}{\partial t}}=\frac{1}{2}M$.
  Injecting into \eqref{eq:11}, this yields $\half
  M=\frac{\kappa}{8\pi}\cA$ as it should.
\item Alternatively, one can compute $\oint_H \cC_\xi$ directly on the
  horizon, or even better, following \cite{Iyer:1994ys}, on the
  bifurcation surface $B$, where $\xi^s=0$, so that now only the first
  term in $\cC_\xi$ will contribute. Indeed, if one chooses the path
  $g^s_{\mu\nu}$ by replacing $M$ through $sM$ in the Schwarzschild
  solution, one has, in Kruskal coordinates,
  $\xi^s=\kappa^s(-U\frac{\partial}{\partial
    U}+V\frac{\partial}{\partial V})$ where $\kappa^s=\frac{1}{4sM}$.
  In these coordinates, the integral $\oint_B C_\xi=\oint_B\int^1_0
  ds\, K^K_{\frac{d\xi^s}{ds}}[g^s]$ becomes $
  \int_0^1 ds\, \frac{4\pi(2Ms)^2}{16\pi} (-\partial_V
  (\frac{d\xi^V}{ds}) + \partial_U(\frac{d\xi^U}{ds}) )=\frac{1}{2}M$,
  as it should.
\end{enumerate}

For the higher dimensional Kerr-AdS black holes discussed below, although
the explicit computations are a bit more involved, the derivation of the
generalized Smarr relation will also follow directly from evaluation of
\eqref{eq:11}.

\mysection{Application to Kerr-AdS black holes}

\subsection[Description of the solutions]{Description of the
  solutions}
\label{sec:descr-solut}

The general Kerr Anti-de Sitter metrics in $n = 2 N +1 + \eps$
dimensions\footnote{In this section we shall use the notations of
  \cite{Gibbons:2004ai} except the spacetime dimension denoted by $n$
  and the indices $a,b$, which run from 1 to $N$, while $i,j$
  run from 1 to $N+\eps$. When $\eps= 1$, $a_{N+\eps} \equiv
  0$.}, where $\eps \equiv n-1 \text{ mod } 2$ were obtained in
\cite{Gibbons:2004uw,Gibbons:2004js}.  They have $N$ independent
rotation parameters $a_a$ in $N$ orthogonal 2-planes.  Gibbons et
al.~start from the $n$ dimensional anti-de Sitter metric in static
coordinates,
\begin{equation}
\bar{ds}^2 = -(1+y^2l^{-2})dt^2 +
\frac{dy^2}{1+y^2l^{-2}}+y^2\sum_{a=1}^N \hat \mu^2_a d\phi_a^2
+y^2\sum_{i=1}^{N+\eps} d\hat \mu^2_i,\label{adsy}
\end{equation}
with $\sum_{i=1}^{N+\eps}\hat \mu_i^2 = 1$. They then consider the
change of variables to Boyer-Linquist spheroidal coordinates
$(\tau,r,\varphi_a,\mu_i)$. These coordinates depend on $N$
arbitrary parameters
$a_a$ and are defined by
\begin{equation}
y^2 \hat \mu_i^2 = \frac{(r^2+a_i^2)}{\Xi_i}\mu_i^2,\qquad \varphi_a =
\phi_a,\qquad \tau = t.\label{coordtra}
\end{equation}
Note that for later convenience, we have renamed the variables
$t,\phi^a$ as $\tau,\varphi^a$ already at this stage. The anti-de
Sitter metric then becomes
\begin{eqnarray}
\bar{ds}^2 &=& - W(1+r^2l^{-2}) d\tau^2 + \frac{U}{V}\, dr^2 +
\sum_{a=1}^N
\frac{r^2+a^2_a}{\Xi_a}\mu_a^2 d{\varphi}_a^2 \label{adstilde}\\
&& \nonumber + \sum_{i=1}^{N+\eps}\frac{r^2+a^2_i}{\Xi_i}d\mu_i^2 -
\frac{l^{-2}}{W(1+r^2l^{-2})}\big( \sum_{i=1}^{N+\eps}
\frac{r^2+a_i^2}{\Xi_i}\mu_i d\mu_i \big)^2,
\end{eqnarray}
where
\begin{eqnarray}
W &\equiv& \sum_{i=1}^{N+\eps} \frac{\mu_i^2}{\Xi_i},\qquad U \equiv
r^\eps \sum_{i=1}^{N+\eps}
\frac{\mu_i^2}{r^2+a^2_i}\prod_{a=1}^N(r^2+a_a^2),\quad
\sum_{i=1}^{N+\eps} \mu_i^2 = 1,\\
V&\equiv& r^{\eps-2}(1+r^2 l^{-2})\prod_{a=1}^N (r^2+a^2_a),\qquad \Xi_i
\equiv 1-a_i^2l^{-2},
\end{eqnarray}
In the coordinates $(\tau,r,\varphi^a,\mu^i)$, the Kerr-AdS solutions
$g_{\mu\nu}$, depending on $N+1$ parameters $M,a_a$, are related to
the AdS metric $\bar g_{\mu\nu}$ as follows:
\begin{eqnarray}
{ds}^2 = \bar{ds}^2|_{\eqref{adstilde}}
+ \frac{2M}{U}\big( W\, d\tau - \sum_{a=1}^N
\frac{a_a \mu_a^2}{\Xi_a}\, d{\varphi}_a \big)^2 + \frac{2M
U}{V(V-2M)}\,dr^2.\label{KerrAdsmetric}
\end{eqnarray}
as can be directly verified by comparing with equation (4.2) of
\cite{Gibbons:2004ai}. In these coordinates, defining the metric
deviations $h_{\mu\nu}$ through
\begin{eqnarray}
ds^2= \bar
{ds}^2|_{\eqref{adstilde}}+h_{\mu\nu}dx^\mu\dx^\nu\label{eq:2}
\end{eqnarray}
and using $U =r^{n-3}+o(r^{n-3})$, $V = r^{n-1}l^{-2}+o(r^{n-1})$, it
is straightforward to see that
\begin{eqnarray}
h_{AB}\sim O(r^{-n+3}), \quad h_{rr} \sim
O(r^{-n-1}),\label{falloff}
\end{eqnarray}
with $A = (\tau,\varphi_a)$, while all other components of
$h_{\mu\nu}$ vanish.

The Killing vectors of the Kerr metric are given in coordinates
$(t,y,\phi_a,\hat\mu_i)$ and $(\tau,r,\varphi_a,\mu_i)$ by
\begin{equation}
k \equiv \Q{}{t} = \frac{\d}{\d \tau},\qquad m^a \equiv \Q{}{\phi_a} =
\Q{}{\varphi_a}. \label{eq:kill2}
\end{equation}

\subsection[Mass and angular momenta]{Mass and angular momenta}
\label{sec:mass-angular-momenta}

\paragraph{Useful integrals}

Let us define the spheroid $S^\infty$ in coordinates
$(\tau,r,\varphi_a,\mu_i)$ by $r=cst \longrightarrow \infty$,
$\tau=cst$. Using $\sqrt{-g}=\sqrt{-\bar g}$ given explicitly in
equation (A.9) of \cite{Gibbons:2004ai} and expressing $\mu_{N+\eps}$
as a function of the remaining $\mu_\alpha$, $1\leq \alpha \leq N+\eps
- 1$, it is straightforward to show that
\begin{eqnarray}
\cA^{sphoid}&\equiv&\int_{S^\infty} \prod_{\alpha=1}^{N+\eps-1}
d\mu_\alpha \prod_{a=1}^{N} d\varphi_a\ \frac{\sqrt{-\bar g}}{r^{n-2}} =
\frac{\cA_{n-2}}{\prod_{a=1}^N \Xi_a},
\end{eqnarray}
where $\cA_{n-2}$ is the volume of the unit $n-2$ sphere, given explicitly
for instance in (4.9) of \cite{Gibbons:2004ai}.

Similarly,
\begin{eqnarray}
\cI &\equiv & \int_{S^\infty} \prod_{\alpha=1}^{N+\eps-1} d\mu_\alpha
\prod_{a=1}^{N} d\varphi_a\ \frac{\sqrt{-\bar g}}{r^{n-2}} \, W =
\frac{2}{n-1}\big( \sum_{a=1}^N \frac{1}{\Xi_a}+ \frac{\eps}{2} \,\big)
\cA^{sphoid}.
\end{eqnarray}
This identity has been verified using {\it Mathematica} up to $n=8$.
We suppose it holds for higher $n$.

\paragraph{Choosing a path}

The path $\gamma$ joining the AdS background to the Kerr-AdS metric is
chosen as follows: for $s\in [0,1]$, $g^s_{\mu\nu}$ is obtained by
replacing $M$ by $sM$ in (\ref{KerrAdsmetric}).

\paragraph{Angular momenta}

Because $m^a=\frac{\partial}{\partial \varphi_a}$ is tangent to
$S^\infty$ and does not vary along the path $\gamma$, one can use (\ref{eq:6})
instead of \eqref{eq:3}
to compute the angular momenta
$\cJ^a$. Explicitly, one gets
\begin{eqnarray}
\cJ^a  &=& \int_{S^\infty} \prod_{\alpha=1}^{N+\eps-1} d\mu_\alpha
\prod_{a=1}^{N} d\varphi_a\ \frac{\sqrt{-\bar g}}{16\pi} \left(
g^{\tau\alpha}g^{rr} g_{\alpha \varphi^a,r}-\bar g^{\tau\alpha}\bar
g^{rr}
\bar g_{\alpha \varphi^a,r}\right)\nonumber\\
&=&\frac{M a_a}{8\pi} (n-1) \int_{S^\infty}\prod_{\alpha=1}^{N+\eps-1}
d\mu_\alpha \prod_{a=1}^{N} d\varphi_a\ \frac{\sqrt{-g}}{r^{n-2}}
\frac{\mu_a^2}{\Xi_a} = \frac{M a_a }{4\pi\Xi_a}\,\cA^{sphoid}.
\end{eqnarray}
Here, the Komar integral evaluated for the background does not contribute
because $\bar g_{\tau \varphi_a}=\bar g^{\tau \varphi_a} = 0$. The
result agrees with the one given in \cite{Gibbons:2004ai}.

\paragraph{Mass}

In order to compute the mass, we evaluate \eqref{eq:3} with
$\xi=k=\frac{\partial}{\partial \tau}$ on $S^\infty$. We have
\begin{eqnarray}
\int_{S^\infty}( -K^K_{k}[g]+K^K_{k}[\bar g])&=&\nonumber \\
\mbox{}\hspace{2cm}&&\hspace{-3cm}\int_{S^\infty}
\prod_{\alpha=1}^{N+\eps-1} d\mu_\alpha \prod_{a=1}^{N} d\varphi_a\
\frac{\sqrt{-\bar g}}{16\pi} \left( g^{\tau\alpha}g^{rr} g_{\alpha \tau,r}
- {\bar g}^{\tau\alpha}{\bar g}^{rr} {\bar g}_{\alpha
\tau,r}\right).\label{eq:int1}
\end{eqnarray}
Let decompose the metric as $g_{\mu\nu} = \bar g_{\mu\nu} +
h_{\mu\nu}$.  The asymptotic behavior~\eqref{falloff} of $h_{\mu\nu}$
implies that $h^\mu_{\,\,\nu} = \bar g^{\mu\alpha}h_{\alpha\nu} =
O(r^{-n+1})$.  Hence, in the expansion of the inverse metric $g^{\mu\nu}$
\begin{equation} g^{\mu\nu} = \bar
  g^{\mu\alpha}(\delta^{\,\,\nu}_{\alpha}
-h^{\,\,\nu}_{\alpha} +
h^{\,\,\beta}_{\alpha}h^{\,\,\nu}_{\beta} -h^{\,\,\gamma}_{\alpha}
 h^{\,\,\beta}_{\gamma} h^{\,\,\nu}_{\beta} +\cdots).\label{decomp}
\end{equation}
only the first two terms will contribute to integral~\eqref{eq:int1},
since the following terms fall off faster and keeping only the first
two terms will give finite contributions, as we will show.  Injecting
this expansion into \eqref{eq:int1}, one gets terms that are at most
quadratic in $h_{\mu\nu}$. The terms of order $0$ will cancel, while
the terms quadratic in $h_{\mu\nu}$ can directly be shown not to
contribute. Hence, only terms linear in $h_{\mu\nu}$ will contribute to
\eqref{eq:int1} with the result
\begin{eqnarray}
\int_{S^\infty} (-K^K_{k}[g]+K^K_{k}[\bar g]) &=& \frac{M}{8\pi}
\int_{S^\infty}\nonumber \prod_{\alpha=1}^{N+\eps-1} d\mu_\alpha
\prod_{a=1}^{N} d\varphi_a\
\frac{\sqrt{-\bar g}}{r^{n-2}} \Big[ (n-1)W-2 \Big] \\
&=&  \frac{M \cA_{n-2}}{4 \pi (\prod_{a} \Xi_a)}\left( \sum_{b=1}^N
\frac{1}{\Xi_b} + \frac{\eps}{2}-1 \right).
\end{eqnarray}

The integral $\oint_{S^{\infty}} \cC_{k;\gamma}$ defined in~(\ref{intcC})
reduces to the integral of the $\xi \cdot \Theta$ part which reads
\begin{equation}
\oint_{S^{\infty}} \cC_{k;\gamma} = \int_0^1 ds\, \int_{S^\infty}
\prod_{\alpha=1}^{N+\eps-1} d\mu_\alpha \prod_{a=1}^{N} d\varphi_a\
\frac{\sqrt{-\bar g}}{16\pi} (D^{(s)}_\sigma h_{(s)}^{r\sigma} -
\partial^r h_{(s)}),\label{intcCKerr}
\end{equation}
where the integrand (but not the integral) depends explicitly on the
path through $g^{(s)}_{\mu\nu}$, $h^{(s)}_{\mu\nu}=\frac{d
  g^{(s)}_{\mu\nu}}{ds}$ (and indices are lowered and raised with
$g^{(s)}_{\mu\nu}$ and its inverse). Note that the equality
$\sqrt{-g^{(s)}} = \sqrt{-\bar g}$ implies $h^{(s)} = 0$. From the
definition of the metric~\eqref{KerrAdsmetric}, one can see that
\begin{equation}
h^{(s)}_{\mu\nu}=h_{\mu\nu}+ o (h_{\mu\nu}),\quad g^{(s)}_{\mu\nu}=\bar
g_{\mu\nu}+ s h_{\mu\nu} + o (h_{\mu\nu}),\label{fall_s}
\end{equation}
where $\bar g_{\mu\nu}$ is the AdS metric and $h_{\mu\nu}$ is defined
in \eqref{eq:2}. Now, as the leading terms in
expression~\eqref{fall_s} give finite contributions to the
integral~\eqref{intcCKerr}, as we will show below, the sub-leading
terms $o(h_{\mu\nu})$ will not contribute. Expanding
$g^{\mu\nu}_{(s)}$ as in \eqref{decomp}, we get
\begin{equation} g_{(s)}^{\mu\nu} \sim \bar
  g^{\mu\alpha}(\delta^{\,\,\nu}_{\alpha}
-s h^{\,\,\nu}_{\alpha} + s^2 h^{\,\,\beta}_{\alpha}h^{\,\,\nu}_{\beta} -
\cdots),\label{decomp2}
\end{equation}
where the indices are raised with $\bar g^{\mu\nu}$ and where $\sim$
indicates that the sub-leading terms in equation~\eqref{fall_s} have
been dropped. Again, we will show below that the first two terms of
\eqref{decomp2} give finite contributions to the
integral~\eqref{intcCKerr}. As the following terms in~\eqref{decomp2}
fall off faster, we can safely ignore them in the computation. If we
now expand the expressions $g_{\mu\nu}^{(s)}$, $g_{(s)}^{\mu\nu}$ and
$h_{\mu\nu}^{(s)}$ in the integrand $\sqrt{-\bar g} D^{(s)}_\sigma
h_{(s)}^{r\sigma}$ in terms of $\bar g_{\mu\nu}$ and of $h_{\mu\nu}$,
we obtain after some work that
\begin{equation}
\sqrt{-\bar g} D^{(s)}_\sigma h_{(s)}^{r\sigma} = \sqrt{-\bar g} {\bar
D}_\sigma h^{r\sigma} + O(r^{-n+1})
\end{equation}
where all the dependence in $s$ appear only in the vanishing term
$O(r^{-n+1})$. Because $h=0$, we have thus shown that $\oint_{S^{\infty}}
\cC_{k;\gamma}$ reduces to the integral over $S^\infty$ of the third term
of \eqref{eq:wald}, with $\xi =k$. Hence, we have shown that at
$S^\infty$, the mass (and, as we have seen before, the angular momenta as
well) can be computed using any one of the equivalent expressions
\eqref{hom}, \eqref{gsuperpot3}, \eqref{anderson}, \eqref{eq:wald} and
\eqref{eq:k0iHamilt} (since $\tau$ is constant on $S^\infty$) of the
linearized theory.

Explicitly, one shows after some computations that $D_\sigma
h^{r\sigma}$ reduces to $r^{-1}h^{rr} + o(r^{-n+2})$.  Therefore,
$\oint_{S^\infty} \cC_{k;\gamma}$ becomes
\begin{eqnarray}
\oint_{S^\infty} \cC_{k;\gamma} &=& \frac{M}{8\pi} \cA^{sphoid}=
\frac{M}{8\pi}\frac{\cA_{n-2}}{(\prod_{a} \Xi_a)}.\label{eq:cC}
\end{eqnarray}
Finally, the energy (\ref{eq:3}) is obtained by summing the two
contributions $\oint(-K^K_k[g]+K^K_k[\bar g])$ and $\oint
\cC_{k;\gamma}$, which gives explicitly
\begin{equation}
\cE  = \frac{M \cA_{n-2}}{4 \pi (\prod_{a} \Xi_a)}\left( \sum_{b=1}^N
\frac{1}{\Xi_b} - \frac{(1-\eps)}{2} \right),\label{eq:ee}
\end{equation}
in agreement with \cite{Gibbons:2004ai,Deruelle:2004mv}.

\subsection[Generalized Smarr relation]{Generalized Smarr relation}
\label{sec:gener-smarr-relat}

We now evaluate the remaining terms in the Smarr
relation~\eqref{PreSmarr}.

The integral $\oint_H K^K_\xi[\bar g]$ evaluated on the surface $r =
r_+$, where the horizon radius $r_+$ is the largest root of $V(r)-2m =
0$, is given by
\begin{eqnarray}
\oint_H K^K_\xi[\bar g] = - \frac{\cA_{n-2}}{8\pi l^2(\prod_a \Xi_a)}
r_+^\eps\prod_{a=1}^N (r_+^2+a^2_a).
\end{eqnarray}
Note that this integral vanishes in Minkowski space $(l\rightarrow
\infty)$.

In Kerr-AdS spacetimes, the Komar integrand $K^K_\xi$ of a
Killing vector $\xi$ is not closed. Indeed, using the equations of motion
$R_{\mu\nu} = - (n-1)l^{-2}g_{\mu\nu}$, we have
\begin{eqnarray}
d K^K_{\xi}[g]&=&\frac{1}{16\pi}(d^{n-1}x)_{\nu}\sqrt{-g} \big(D_\mu
D^\mu\xi^\nu- D_\mu D^\nu\xi^\mu \big)\\
&=& -\frac{n-1}{8\pi l^2}(d^{n-1}x)_{\nu}\sqrt{-g} \xi^\nu.
\end{eqnarray}
Because $\sqrt{-g}=\sqrt{-\bar g}$, we have
$d(-K^K_{\xi}[g]+K^K_{\xi}[\bar g])=0$. It then follows from the
definition of $\cC_{\xi;\gamma}$ (see equation \eqref{eq:3}) and from
the identity $dK_{\xi;\gamma}=0$ that $d\,\cC_{\xi;\gamma} = 0$. We
thus can move the integral on the horizon back out to infinity,
\begin{eqnarray}
\oint_H\cC_{\xi;\gamma} =\oint_{S^\infty}\cC_{k;\gamma}
+\Omega_a\oint_{S^\infty}\cC_{m^a;\gamma}.
\end{eqnarray}
The first term on the right hand side has already been computed in
\eqref{eq:cC}, while the second term vanishes because
$m^a=\frac{\partial}{\partial \varphi^a}$ does not vary along the path
and is tangent to $S^\infty$.

We can now write the Smarr formula (\ref{PreSmarr}) as
\begin{eqnarray}
\cE - \Omega_a \cJ^a &=& \frac{\kappa \cA^{sphoid}}{8\pi} +
\frac{\cA^{sphoid}}{8\pi}\big( M - \frac{r_+^\eps}{l^2}\prod_{b=1}^N
(r_+^2+a^2_b) \big),\label{eq:smarrfin}
\end{eqnarray}
in complete agreement with the results obtained by Euclidean methods in
\cite{Gibbons:2004ai}.

In the limit $l \rightarrow \infty$, we recover the Smarr formula for Kerr
black holes in flat backgrounds since then $\cA^{sphoid} = \cA_{n-2}$,
$\oint_H K^K_\xi[\bar g]= 0$. Combining \eqref{eq:cC} with \eqref{eq:ee}
then gives $\oint \cC_\xi = (n-2)^{-1}\cE$. Injected into
\eqref{PreSmarr}, we finally have
\begin{eqnarray}
  \frac{n-3}{n-2}\cE - \Omega_a \cJ^a &=& \frac{\kappa \cA_{n-2}}{8\pi}.
\end{eqnarray}

\appendix

\mysection{Integrability of charges for trivial de Rham cohomology}

In this appendix, we follow the notation of \cite{Barnich:2001jy},
appendix A. More details can be found in \cite{Andersonbook}. Note
however that in the considerations below we are considering jet spaces
where the fibers include not only the fields $\phi^i$ and their
derivatives, but also the gauge parameters $f^\alpha$. We will use the
notation $\phi^\Delta\equiv (\phi^i,f^\alpha)$.

In particular, $S=\int d^nx\, L$ is the action of a local gauge field
theory (e.g., the Einstein-Hilbert action), $\phi^i$ are the fields (e.g.,
$g_{\mu\nu}$), $f^\alpha$ the gauge parameters (e.g., $\xi^\mu$) and
$R^i_{\alpha}(f^\alpha)$ denote a generating set of gauge transformations
(e.g., $\cL_\xi g_{\mu\nu}$). The weakly vanishing Noether current $S_f$
is defined through \bea \vdd{L}{\phi^i}R^i_\alpha(f^{\alpha})d^nx=d_H
S_{f},\ S_{f}\equiv S^i_\alpha(\vdd{L}{\phi^i},f^{\alpha}).
\label{full}\eea

If $\omega^{p,s}$ is a $p$ form in $dx^\mu$ and an $s$ form in
$d_V\phi^a_{(\lambda)}$, the horizontal homotopy operator (see eq
(4.13) of \cite{Andersonbook}) is defined by \bea
h^{p,s}_H\omega^{p,s}=\frac{1}{s}
\frac{|\lambda|+1}{n-p+|\lambda|+1}\partial_{(\lambda)}
[d_V\phi^\Delta\frac{\delta}{\delta d_V
  \phi^\Delta_{(\lambda)\rho}}\frac{\partial\omega^{p,s}}{\partial
  dx^\rho}], \eea where a sum over the multi-index $(\lambda)$ is
understood.  If \bea k_{f}[d_V\phi;\phi]=-h^{n-1,1}_Hd_V
S_{f},\label{7} \eea it follows from the homotopy formula
$\omega^{p,s}=d_H h^{p,s}_H\omega^{p,s}+h^{p+1,s}_Hd_H\omega^{p,s}$,
valid for $s\geq 1$, that \bea d_Hk_{f}[d_V\phi;\phi] =-d_V S_{f
}-h^{n,1}_H d_V\big(R^i_\alpha(f^\alpha)\frac{\delta
  L}{\delta\phi^i}\big)d^nx.
\label{8} \eea This implies that sufficient conditions for the right hand
side to vanish are that $\phi^i$ satisfies the field equations,
$\frac{\delta L}{\delta\phi^i}|_{\phi}=0$, $\delta\phi^i$ satisfies
the linearized field equations around $\phi^i$, $(d_V\frac{\delta
  L}{\delta\phi^i})|_{\delta\phi,\phi}=0$, and $f^\alpha$ is a
reducibility parameter for $\phi^i$, $R^i_\alpha[\phi]( f^\alpha)=0$
(which, in the case of relativity, translates into the requirement
that $\xi^\mu$ is a Killing vector).

Note that $k_{f}[d_V\phi;\phi]$ defined in equation \eqref{7} differs
from the $k_{f}[d_V\phi;\phi]$ defined by the homotopy formula
involving the $\phi^i$ alone (which is the one explicitly used in
subsection \bref{covexp}). The additional piece will however be linear
and homogeneous in $\frac{\delta L}{\delta\phi^i}$ and its
derivatives, hence it vanishes when evaluated for solutions of the
field equations and can be safely dropped for the computation of the
charges. Furthermore, it will also not affect the considerations
below.

Applying $d_V$ to this equation gives
\bea d_Hd_V k_f=d_Vh^{n,1}_H
 d_V\big(R^i_\alpha(f^\alpha)\frac{\delta L}{\delta\phi^i}\big)d^nx.
\label{6}\eea
It follows that sufficient conditions for the right hand side to
vanish are now that $\frac{\delta L}{\delta\phi^i}$,
$d_V\frac{\delta L}{\delta\phi^i}$, $R^i_\alpha(f^\alpha)$ and
$d_V[R^i_\alpha(f^\alpha)]$ should vanish when evaluated for
$\phi^i,f^\alpha,\delta\phi^i,\delta f^\alpha$. This proves the first
claim in subsection \bref{pathind}.

For the second claim, we first note that the vertical homotopy is given by
\bea
h_V\omega^{p,s}=\int^1_0\frac{dt}{t}\big(\phi^\Delta_{(\mu)}
\frac{\partial\omega^{p,s}}{\partial
  d_V\phi^\Delta_{(\mu)}}\big)[x,dx,t\phi,td_V\phi],
\eea
with
\bea
\omega^{p,s}-\omega^{p,s}[x,dx,\phi=0,d_V\phi=0]=d_V
h_V\omega^{p,s}+h_Vd_V\omega^{p,s},
\eea and (see e.g.  (4.11) of
\cite{Andersonbook}) \bea d_Hh_V\omega^{p,s}=-h_Vd_H\omega^{p,s}.
\label{comm} \eea Assume now that $k_f[d_V\phi;\phi]$ is evaluated for a
family of solutions $\phi^i_a$ and associated reducibility parameters
$f^\alpha_a$ depending on continuous parameters $a^A$.
In this case, the solution space can be
identified with the parameter space. The differential $d_V$ then reduces
to the exterior derivative in parameter space, $d_a= da^A \frac{\partial
}{\partial a^A}$ and the vertical homotopy $h_V$ reduces to the standard
de Rham homotopy operator $h_a$ in parameter space. In particular, the
conditions discussed below equation \eqref{6} then hold. It follows from
\eqref{6} that $d_H d_a k_{f_a}[d_a\phi_a;\phi_a]=0$, where
$d_H=dx^\mu\frac{d}{dx^\mu}$ reduces to the total derivative with respect
to $x^\mu$. Because $d_a k_{f_a}[d_a\phi_a;\phi_a]$ is a closed $n-2$ form
in $dx^\mu$ depending on parameters $a^A,da^A$, it follows, because of the
assumption that there is no cohomology in degree $n-2$ in $x^\mu$ space,
that
\begin{eqnarray}
d_a k_{f_a}[d_a\phi_a;\phi_a]=d_H l(a,da,x,dx),\label{eq:21}
\end{eqnarray}
where $l(a,da,x,dx)$ is some $n-3$ form in $dx^\mu$ depending on the
parameters $a^A,da^A$. (Note that $l(a,da,x,dx)$ is not necessarily
built out of the fields and a finite number of their derivatives).
Integrating over a closed $n-2$ dimensional surface $S$, we get $d_a
\oint_S k_{f_a}[d_a\phi_a;\phi_a]=0$. If there is no de Rham
cohomology in degree $2$ in parameter space, this implies $\oint_S
k_{f_a}[d_a\phi_a;\phi_a]=d_a Q(a)$ for some function $Q(a)$. More
precisely, in the case where the parameter space is $\mathbb R^n$, one
can apply the homotopy operator $h_a$ and \eqref{comm} implies that
$k_{f_a}[d_a\phi_a;\phi_a]=-d_Hh_a l(a,da,x,dx)+d_a h_a
k_{f_a}[d_a\phi_a;\phi_a]$.  We thus find that $Q(a)=\oint_S h_a
k_{f_a}[d_a\phi_a;\phi_a]$. (Again $h_a k_{f_a}[d_a\phi_a;\phi_a]$ is
not necessarily built out of the fields and a finite number of their
derivatives.) 

\section*{Acknowledgments}
The authors are grateful to M.~Henneaux, J. Zanelli, R. Troncoso and
C. Mart\'{\i}nez for useful discussions. This work is supported in
part by a ``P{\^o}le d'Attraction Interuniversitaire'' (Belgium), by
IISN-Belgium, convention 4.4505.86, by the National Fund for
Scientific Research (FNRS Belgium), by Proyectos FONDECYT 1970151 and
7960001 (Chile) and by the European Commission program
MRTN-CT-2004-005104, in which the authors are associated to
V.U.~Brussel.

\providecommand{\href}[2]{#2}\begingroup\raggedright\endgroup

%\bibliography{E:/Physics/Bibliography/master2}
%\bibliography{master_Biblio}

\end{document}